# Information generating, sharing and manipulating Source-Reservoir-Sink model of self-organizing dissipative structures

Shoaib Ahmad

National Center for Physics, QAU Campus, Shahdara Valley, Islamabad 44000, Pakistan

sahmad.ncp@gmail.com

## Abstract

Shannon's information-theoretic description of the signal transmitter, the channel and receiver is extended to the network of self-organizing dissipative structures consisting of a source, a reservoir and a sink. The information-generation by the source is subjected to controlled manipulation by the reservoir before being transmitted to the sink. The reservoir can have memory and variable capacity for information storage. The reservoir can be physical and tangible like a lake or virtual like Google. In temporal terms, the reservoir as a manipulator of the received information, could last for millions of seconds or billionth of a second. The role of the reservoir in building the manipulative capacity for information storage and selective sharing is illustrated by the characteristic of asymmetric exchange between the reservoir and the sink. A Box-model is used to develop the model to represent material, process and information sharing among the source, the reservoir and the sink. The number of boxes is varied and the diagnostic tools like relative entropy, entropic cost of the output and fractal dimension are evaluated to characterize the model. The model is applied to self-organizing carbon cages with the end-directed evolution of the Buckyball.

**Self-organizing dynamical systems are treated analogous to the networks of information-generating and sharing components. The Source is the generator and initiator of all information. This information is transmitted towards the eventual receiver–the Sink, through the intermediate stage of the Reservoir. Reservoir is treated as the only link between the Source and the Sink. An information-theoretic Source-Reservoir-Sink (SRS) model is developed where Reservoir is analogous to Shannon's Channel. Reservoir is shown to be the repository of material, energy and the consequent information. In Reservoir, material or information received from the Source can be manipulated. This information is delivered to the Sink. The Sink may receive (i) material output in various shapes, configurations or forms, or (ii) signals that are representative of the processes that occurred in the Reservoir, or (iii) useable or waste energy, etc. The Sink is the recipient of the material, information or it could also participate in the dynamical processes. The interactive participation of the Reservoir with Source and Sink is the core of the SRS model. Reservoir is shown to consist of one or more steps or stages of the self-organizing dissipative structures. SRS model is developed through interconnected Boxes that share material and information. Probability distributions are constructed that are used to calculate Shannon entropy, relative entropies and fractal dimensions. The model is applied to provide the information-theoretic description of the ensembles of self-organizing carbon cages with the emergence of the perfectly symmetrical $C_{60}$ cage. The model has already been applied to fragmenting nanotubes.**

# I. INTRODUCTION

Emergence of self-organizing dissipative structures in a variety of energy consuming, entropy generating open systems has been described by a large number of eminent scholars, only a brief selection of the few references, relevant to the work described here, is given as refs. [1-10] and the references therein. This communication has emerged out of the investigations, experimental and theoretical, on various mechanisms of formation and fragmentation of carbon nanostructures. The mechanisms of formation of fullerenes in the soot generated by energetic heavy ion bombardments of graphite [11-13] and in the regenerative soot of hollow graphite discharges [14-17] have intrinsic similarities with the processes of fragmentation of the irradiated $C_{60}$-fullerite [18] and carbon nanotubes [19]. With the help of the model developed here, we show that the energy dissipation and entropy generation during the formation and fragmentation stages can be described by the information-theoretic tools [20-26]. Two persistent themes have been noticed; (a) Maximum Entropy Production [27-29] may not be associated with the self-organizing, emergent structures, and (b) the inter- and intra-structural proesses play significant roles. After developing the general principles of the Source-Reservoir-Sink (SRS) model, we focus on the emergence of the most symmetric carbon cage-the Buckyball [30]. It emerges out of the self-organizing soot that contains carbon atoms, molecules, clusters and closed cages. The closed cages initially are formed in all shapes, sizes and isomeric densities. The cage transformations through fragmenting sequences and intra-cage bond reorientations [31] produce the icosahedral $C_{60}$. It has been described as an 'order out of disorder' transition of self-organizing dissipative structures [32]. Kinetic and thermodynamic arguments to describe this process have had varying degrees of success [33-37]. In this communication, the far from equilibrium, irreversible dissipative structures are monitored through their information generation during the spatial and temporal stages, mostly within the Reservoir of the SRS model.

Shannon entropy is evaluated for each constituent of the irreversible dissipative structure that participate in the formative or fragmenting processes. Our model has the same basic design as that of Shannon's three-component system that is composed of signal Transmitter, the Channel and the signal Receiver. The Source symbolizes the Transmitter, Resevoir is equivalent to the Channel and Sink is the Receiver in our SRS model. Three significant modes of the model' application are (a) free-flow, (b) source-friendly and (c) the reservoir-specific modes are identified. The source is the fountainhead where the process begins with the generation of the material and the information. The reservoir receives the material and information from the source at certain rates. Depending upon the design, the style, capacity and modes of communication with the sink, the material and information-sharing between the reservoir of information and the sink or the receiver occurs. The sink, in our model, is the sole recipient of the transfer of material, the associated processes and information from the reservoir. It is the end-directed evolution of the dissipative dynamical system of SRS. The transfer rate is determined by the mutual inter-connectivity and the dynamical system-specific parameters. We will first present the general features of the SRS model and then in the second half of the paper, a specific application of the model will be discussed. The basic difference between our model and Shannon's description of a Source–Channel–Receiver is that our reservoir's ability of manipulation, enhancement or degradation of the information (or material) received from the source is a desireable feature unlike the noise generated by the Channel in Shannon model. Furthermore, the reservoir can have memory, fixed or variable. The reservoir may also bilaterally interact and exchange information (or material) with the sink. Our description of such a dissipative structure of source-reservoir-sink can display self-organisational character. The information-theoretic parameters are derived from Shannon entropy calculated for each constituent and the relative entropies between the constituents of the SRS model. We derive the entropic cost of the evolution of the emergence as a diagnostic tool. The information or fractal dimension of all SRS constituents categorizes the dissipative character of the end-directed evolution [38].

## II. THE MODEL

The SRS model is an effort to start from the basics and build the foundation for understanding the self-organizational routes undertaken by certain dissipative structures. Our focus is the emergence of order out of disorder. The reverse process of the disorder generated in ordered structures like the single-walled carbon nanotubes has recently been discussed with similar information-theoretic arguments elsewhere [26]. After providing a general description of the SRS model, we will discuss the evolution of ensembles of carbon cages out of the disorder of hot carbon vapor. The model will be used to describe the emergence of the most symmetric cage-Buckyball from ensembles of fullerenes of all sorts of shapes, sizes and isomeric variations. In this case, the icosahedral $C_{60}$ is the end-directed evolution of the self-organizing, dissipative dynamical system.

The first assumption of the model is that the flow or the exchange of material between the source reservoir and the sink can be discretized as pulses defined by the exchange rates. For example, in the case of two boxes connected by a controllable valve, the material can be transferred at a well-defined rate. If box#1 had 512 marbles or 512 liters of liquid, it can be exchanged with the box#0 at the rate of 1/2 by sharing 256, 128, 64, 32, 16, 8, 4, 2 and 1 in ten steps. At this stage, the time duration between pulses are not considered. The transferred quantity is the main parameter.

The probability paradigm is at the heart of the model. It defines the dissipative structure, the relative performance of its interactive constituents and displays the emergent characteristics of the modified or new structures. The normalized probability mass distributions for the source, reservoir and the sink are evaluated for different rates of flow and for various combinations thereof.

The probability for a set of distributive stages $\zeta$ is referred to as $p(\zeta)$. The uncertainty is defined and measured as $\ln(1/p(\zeta))$. This function has special characteristics and has also been named 'surprise'. Hence the Kolmogrov measure of uncertainity [23]

$$i_\zeta = \ln(1/p(\zeta)) \qquad (1).$$

The special characteristics associated with $i_\zeta$ are related with the fact that at low $p(\zeta)$ its value increases. The sharply increasing rate of $i_\zeta$ has the surprise [22]. We will illustrate this aspect in the section III.

Entropy can be evaluated from the product $p(\zeta)\ln(1/p(\zeta))$. The sum over all $\zeta$ of this product is the well-known Shannon entropy or the information I

$$I = \sum_\zeta p(\zeta)\ln(1/p(\zeta)) \qquad (2).$$

It must be pointed out that $i_\zeta$ and $I$ are dependent upon the measure $\zeta$ and hence the flow rate. Different configurations of the system (SRS) will yield varying $i_\zeta$ and $I$ corresponding to the various rates of flow and the increasing or decreasing number of distributive stages $\zeta$.

Relative entropy is calculated for the dynamical systems to provide a measure of the Kuulback-Leibler [39] distance between two probability measures like $p_x \equiv p_{source}(\zeta)$ and $p_y \equiv p_{sink}(\zeta)$ which are the probability distributions of the source and the sink, respectively. Relative probability is evaluated as

$$D(p_x \parallel p_y) = \sum p_x(\zeta) \ln(p_x(\zeta)/p_y(\zeta)) \qquad (3).$$

The fourth diagnostic tool is the fractal dimension. The summation over every sequence of $\zeta$ steps or stages, generates information $I_i \equiv \sum_\zeta p(\zeta)\ln(1/p(\zeta))$, is the net information generated by the probability distribution $p(\zeta) \equiv p_i(\zeta)$; where $i = x, y$ for respective probability distributions of dource, reservoir or the sink.

Fractal analysis based on $I$, for various flow rate configurations will be shown to be an indicator of the impact of the different phases or the modes of flow of the dissipative structures. Fractal or the information dimension is defined as [40-42]

$$d_I^x = \sum_\zeta p_x(\zeta) \ln(1/p_x(\zeta))/(ln(1/\zeta)) = I/\ln(1/\zeta) \qquad (4).$$

In equation (4), $\zeta$ is the number of distributive stages or the measure required to obtain information. The fractal dimension can be of the reservoir, the source or the sink; similar notation is used for all three components of SRS model.

An information-theoretic parameter is defined to provide an entropic cost measure of the emergence of the end-directed dissipative structure. It is the ratio of the total information generated by all the constituents $\sum I_x$ to that generated by the emergent one $I_e$ as

$$i_e^\Sigma = \sum I_x/I_e \qquad (5).$$

## III. FREE-FLOW MODE: SOURCE AND SINK ONLY

Let us consider a two-component model of a source and a sink, without the reservoit and hence no channel in Shannon's description of an information-network. In this, simplified Source and the Sink model, we ignore the material or the information lost in the transfer. The 2-Box version of the model assumes that the information generated by the Source or Box#1 is received at the pre-determined rate by the Box#0, without loss, manipulation of modification. We intend to develop the initial, essential information theoretic parameters that are cruicial for the understanding of the role of the reservoir in detail, that will be introduced in the next section. It will be shown that the Reservoir emerges as the manipulator of the original signal or the initial impetus given to the dynamical system.

Let us look at the probability distribution of the quantity of material shared between the source and the sink in Figure 1. With a fixed amount of material, the Source empties and the Sink being the sole recipient receives. The emptying of the Source and filling of the Sinks occurs at a pre-determined rate. The starting probabilities $p_i$ for the Source and the Sink are 1 and 0 at $\zeta = 0$. The respective final probabilities $p_f$ for the two are ~0 and ~1 as $\zeta \to max$.

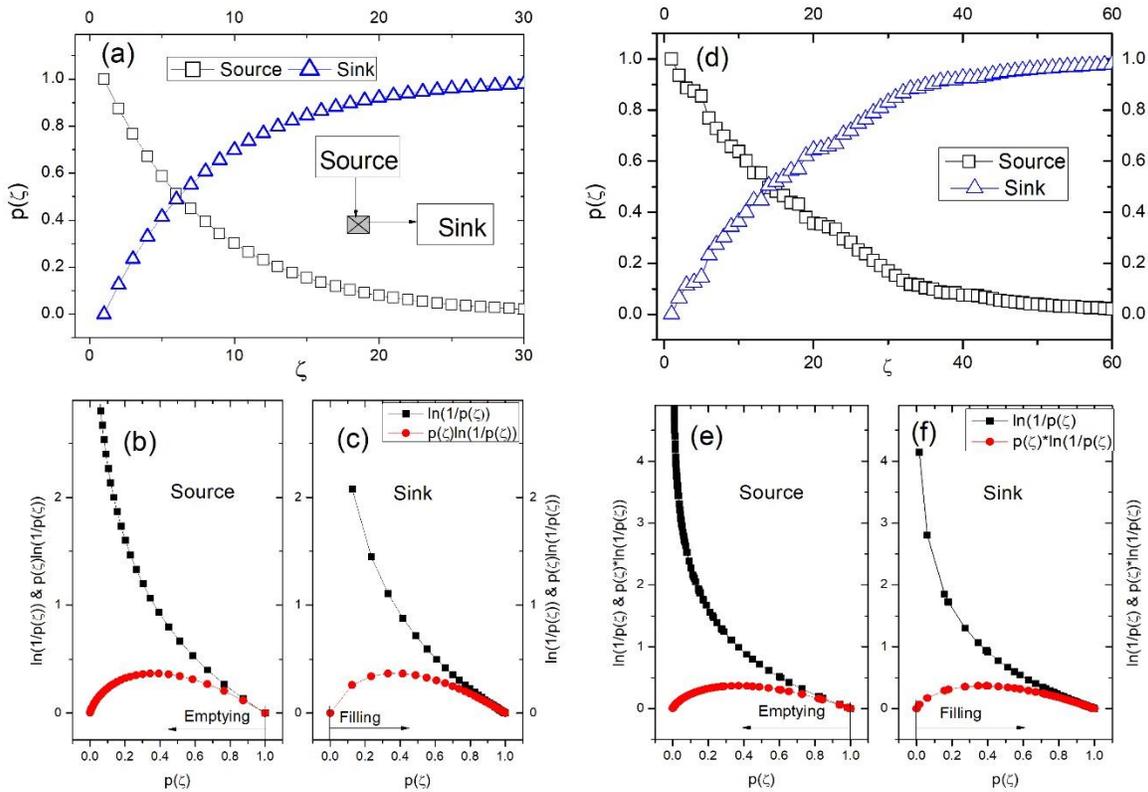

FIG. 1. (a) The two boxes Source and Sink are shown, connected with a controllable valve that regulates and times the flow of material from Source to the Sink. The probability $p(\zeta)$ is plotted for the two as a function of the number of $\zeta$-times the flow occurred. In the figure the rate of flow is set at 1/8th of the total (or the remaining) material in the Source. (b) shows the profile of emptying of Source in terms of the uncertainty $\ln(1/p(\zeta))$ and entropy $p(\zeta)\ln(1/p(\zeta))$. (c) has the same two parameters for the Sink that is filling. The arrows show the directions of emptying

and filling. (d) A random number generated between 0 and 1 is multiplied by 1/8 determines the flow from Source to the Sink. The emptying/filling $\zeta$ range is extended to about twice as compared to that of figure 1. The randomness of the choice of the flow rate is reflected in the $\ln(1/p(\zeta))$ and entropy graphs in (e) and (f).

Figure 1(a) is based on the iterations with constant fixed flow @1/8. The probability distributions and the profiles of emptying and filling are symmetric. The initial, starting probabilities are 1 for the Source and 0 for the Sink indicated by arrows in the figure. For a random element introduced into the flow rate, one can see a different flow profile as shown in figure 1(d). The rate of flow at every step is iterated with a random number chosen between 0 and 1. This is multiplied by 1/8 and the resultant flow is determined by a random process. The maximum rate of flow is 1/8th for the material remaining in the Source.

The importance of the plots of $\ln(1/p(\zeta))$ versus $p(\zeta)$ in Figs. 1(b), (c), (e) and (f) is due to their high values near low $p(\zeta)$. It is a measure of the uncertainty and in the case of the Source which is emptying, the surprise occurs when the material left gradually reduces to nothing. For the Sink, the value of $\ln(1/p(\zeta))$ near low probability values is not as drastically high as it is for the Source. It is due to the state of the Sink and the associated information. There is lesser surprise in 1(c) and 1(f) when more material is received. The simple 2-Box model displays the essential features of a connected, information-sharing system with inter-dependent probability distribution functions, without the intervening reservoir.

## IV. THE SOURCE-RESERVOIR-SINK (SRS) MODEL

If a third box is added between the source and the sink, then a completely different flow dynamics emerges. Figure 2 shows the probabilities of the quantities of the material for the rate of flow @1/2 from the Source and @1/2 from the Reservoir to the Sink. These are the similar rates that were used in Figure 1 for the 2-Box model.

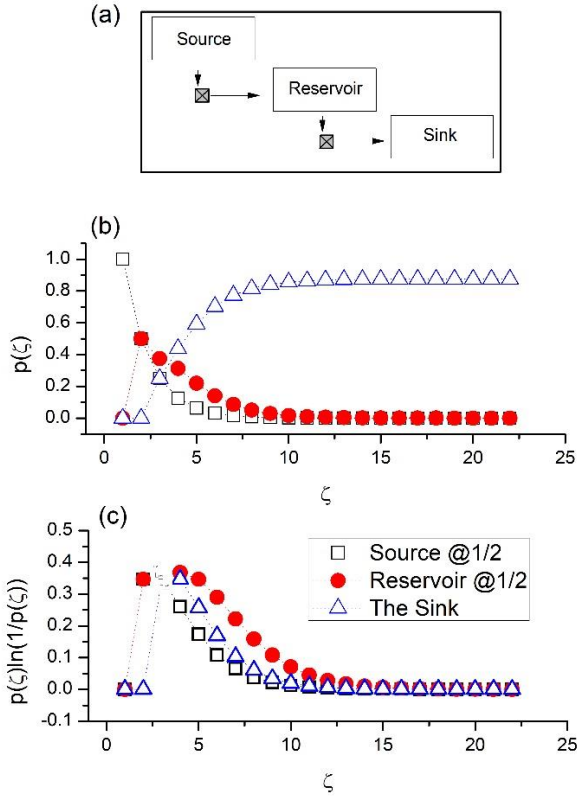

FIG. 2. (a) Schemetic diagram of the Source-Reservoir-Sink connected by 2 valves that can regulate the flow, independently or synchronously. (b) The probabilities $p(\zeta)$ as a function of $\zeta$ are plotted for the three boxes representing the Source, Reservoir and the Sink. (c) Plots of $p(\zeta)\ln(1/p(\zeta))$ versus $\zeta$ shows the entropic profiles as a function of the material/information transfer stages $\zeta$.

Let us now discuss the role of the intermediate stage between the Source where the information or the original signal is created and the Sink receives it in modified form after passing through the Reservoir. In information theory, the Channel can be the source of noise, degradation and alteration of the original signal. In our model too, the Channel or the Resevoir plays the similar, but desireable, manipulable role. It is the repository of information, material or ideas. Some of its features may need to be controlled for desirable effects. We investigate these aspects where a dissipative dynamical system self-organizes and new structures emerge out of the interactions amongst the ingredients of the material received in the reservoir. An end-directed evolution of an emergent structure may occur that can be categorized using the diagnostic tools discussed in section I.

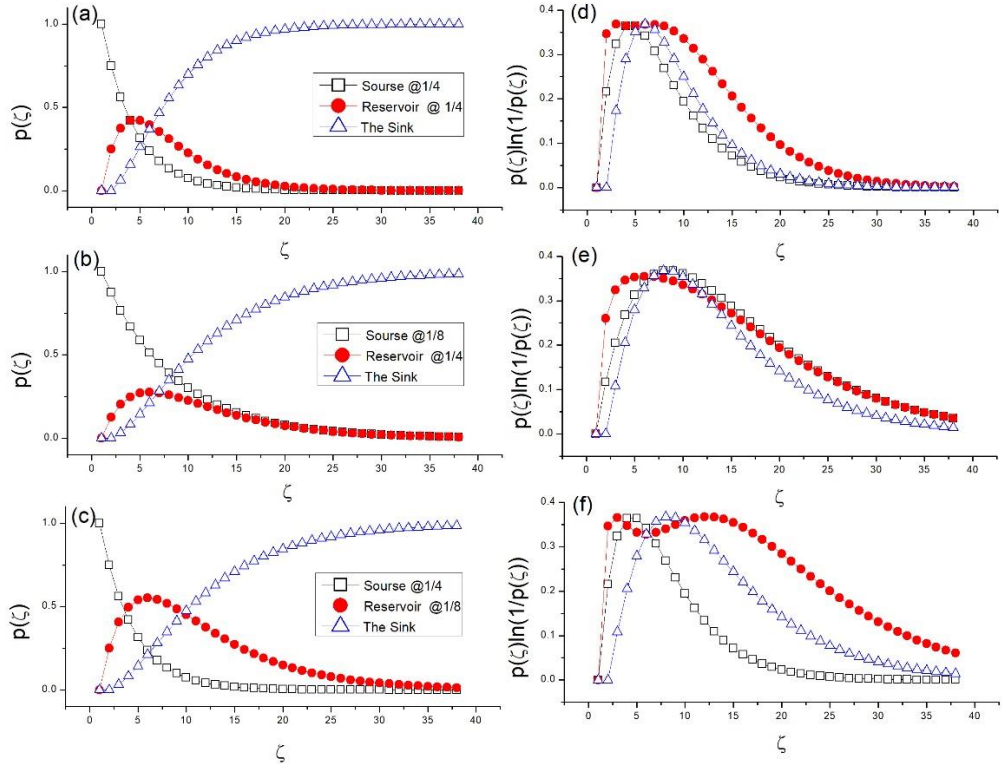

FIG. 3. Varying rates of the input and discharge by the reservoir show the impact on the retention capacity. Probabilities $p(\zeta)$ are plotted as a function of $\zeta$ in (a)–(c). (a) The Source and the Reservoir discharge material @1/4. (b) Higher retention occurs for input @1/4 and discharge @1/8th to the Sink. Entropic profiles $p(\zeta)\ln(1/p(\zeta))$ are plotted for the Source, Reservoir and the Sink in (d), (e) and (f).

The mode shown in (b) is Source-friendly while the one shown in (c) is Reservoir-friendly as the Reservoir empties at a lower rate as compared with that of the Source. Faster rates of the arrival of material coupled with the slower rates of discharge can be seen to build the capacity of the Reservoir as is evident from the comparison of 3(d), 3(e) and 3(f).

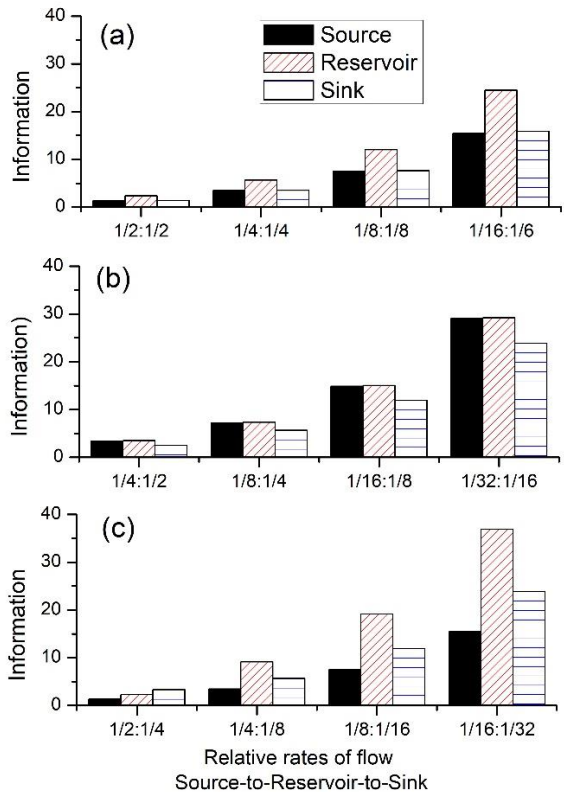

FIG. 4. The impact of different input and output rates of the reservoir are shown on the three modes of information generation and sharing. (a) The free-flow regime of equal rates of input and output. The reservoir builds capacity at the expense of the source and sink. (b) shows slower inputs and higher output rates. (c) The reservoir-friendly mode with higher rates of input and lower rates of output. The possibility of information manipulation by the reservoir increases in this case.

Figure 4 shows the effect of varying the rates of flow of material or information between the Source-to-Reservoir and Reservoir-to-Sink. Information generated by the flows obtained for three different combinations of flow. In figure 4(a) the histograms of equal rates of flow $R_{2\to1}=R_{1\to0}$ for four different rate combinations 1/2:1/2 up to 1/16:1/16 are plotted. Almost equal amount of information is generated by the Source and the Sink with the Reservoir capacity of information building up $I_{Source} \approx I_{Sink} < I_{Reservoir}$. Figure 4(b) has slower rate of emptying of the Source as compared with that of the Sink. The two rates are such that $R_{2\to1}<R_{1\to0}$. In this case the information generated by the Source and Reservoir are of similar order of magnitude and $I_{Source} \approx I_{Reservoir} > I_{Sink}$. Figure 4(c) demonstrates that for faster arrival and slower out-flow from Reservoir, the Reservoir retains higher levels of material the associated information $I_{Reservoir} > I_{Sink} > I_{Source}$.

**Table I.** The composition of relative entropies between the probability distributions of the Source and the Sink and the reverse between the Sink and the Source, for three different sets of flow rates; equal rates $R_{2\to1}=R_{1\to0}$, $R_{2\to1}<R_{1\to0}$ and $R_{2\to1}>R_{1\to0}$. Next two columns have the ratio of the two relative entropies, $D(p_2 \parallel p_0)$, $D(p_0 \parallel p_2)$. The last column has the entropic cost parameter $i_e^\Sigma$.

| Flow rates $R_{2\to1}: R_{2\to1}$ | $D(p_2 \parallel p_0)$ | $D(p_0 \parallel p_2)$ | $i_e^\Sigma$ |
|---|---|---|---|
| 1/2:1/2 | 98.43 | 155.55 | 3.76 |
| 1/4:1/4 | 123.04 | 184.34 | 3.59 |
| 1/8:1/8 | 264.35 | 376.1 | 3.55 |
| 1/16:1/16 | 356.3 | 502.46 | 3.52 |
| 1/4:1/2 | 33.44 | 126.54 | 3.79 |
| 1/8:1/4 | 70.57 | 163.67 | 3.22 |
| 1/16:1/8 | 154.1 | 336.24 | 3.23 |
| 1/32:1/16 | 218.34 | 432.05 | 3.19 |
| 1/2:1/4 | 48.4 | 55.66 | 3.72 |
| 1/4:1/8 | 64.42 | 73.31 | 3.55 |
| 1/8:1/16 | 142.26 | 159.14 | 3.51 |
| 1/16:1/32 | 196.32 | 211.22 | 3.44 |

The relative entropies of two probability distributions that are connected by intermediary stage of a Reservoir, defined as the Kulbach-Leibler distance. In our case we denote these distributions as $p_2 \equiv p_{Source}$ and $p_0 \equiv p_{Sink}$. The relative entropy between the Source and Sink, from equation (3) $D(p_2 \parallel p_0) = \sum p_2(\zeta) ln(p_2(\zeta)/p_0(\zeta))$ and the relative entropy of the distribution of the Sink with the Source is $D(p_0 \parallel p_2) = \sum p_0(\zeta) ln(p_0(\zeta)/p_2(\zeta))$. These have been tabulated in Table I. It is noticeable that all combinations have almost similar entropic cost for the information transfer. Depending upon the physical constitution, conditions and the environment of the dissipative dynamical system, any of the above three requirements may be suitable for the manipulation of the information by the Reservoir. Reservoir, therefore, is the linkpin of the dynamical system. Reservoir in the model receives, retains and regulates the information. The two relative entropies are not equal $D(p_2 \parallel p_0) < D(p_0 \parallel p_2)$ for all combinations of flow rates. The inherent asymmetry is visible. The relative entropy of Sink with respect to Source is bound to be higher due to the nature and content of the flow towards Sink that is manipulated by Reservoir. Reservoir is the mediator and modulator of information transfer between Source and Sink. Relative entropy is a measure of inefficiency of determining the true probability distribution. This aspect is highlighted in Table I. Only in the case of the 2-Box model $D(p_1 \parallel p_0) < D(p_0 \parallel p_1)$; due to the absence of the reservoir.

## V. THE CASE OF MULTIPLE SOURCES AND THE EXTENDED RESERVOIR

In the preceding section, the essential characteristics of an idealized SRS model based on 3 boxes, each representing one of the constituents were presented. In real physical situations, there can be more than one sources and reservoirs. Even the simplest environmental example of a lake feeding on waters from the surrounding hills and draining to a river, contains multiple sources of water. There can also be more than one lakes and rivers. One needs to deal with multiple Sources and determine how the inter-Source links are established. The SRS character of the Multiple Sources configuration enhances the role and the performance of the Reservoir. Except the first source and the Sink, all other intervening sources become the extended reservoir with greater information-manipulating capabilities. Continuing with the notation of boxes, we extend the number of boxes first to six and later to eleven; each sharing its contents, at a pre-determined rate with the next box. In such a case, all, except the first and the last one, bilaterally share. These collectively act as the reservoirs in this example. The model with six-boxes will be discussed in this section. There is a direct implication of the six-box scenario with the real world's physical example discussed in section VIII where the emergence of the Buckyball out of the fragmenting ensembles of fullerenes will be shown to be directly relevant. In section VIII, the case of the ensemble of carbon cages $C_x$ represented by $\sum_{60}^{70} C_x$ will be considered. It contains the set of six fullerenes $C_{60}$, $C_{62}$, $C_{64}$, $C_{66}$, $C_{68}$ and $C_{70}$. The sequence of fragmenting of larger cages into the next smaller cage and a $C_2$ molecule, will be discussed by using the information-theroretic tools developed in the previous sections. The role of the Reservoir will be elaborated to describe the cage-to-cage transformations.

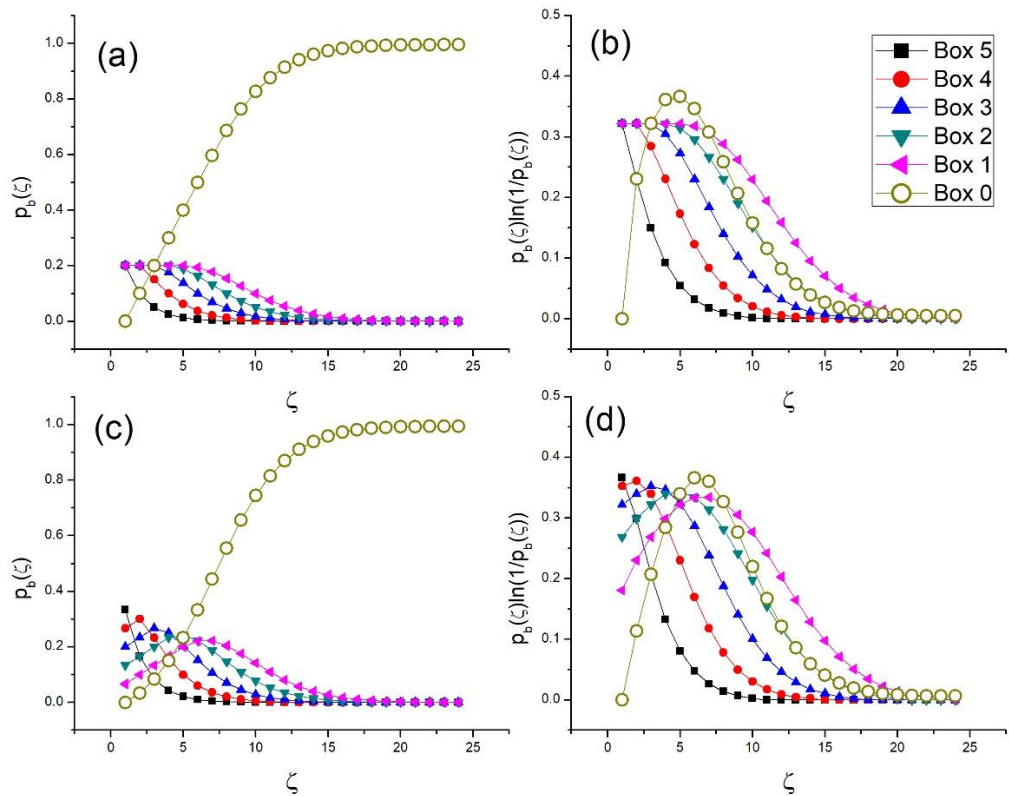

FIG. 5. (a) The probability and (b) entropy profile for a 6-Box system where box#5 to #1 have equal amount of material. Box#0 has none to start with. (c) Probability profiles as a function of $\zeta$ when there are unequal quantities in Box#5 to Box#1 in the ratio 1:0.8:0.6:0.4:0.2 and (d) the associated entropic profiles.

The 6-Box setup with interconnecting valves is similar to the 3-Box arrangement shown in figure 2(a). In the results shown in figure 5, Box#5 to Box#1 all can initially be taken as the Sources that release their contents @1/2 to the next one. Amongst these the Box#4 to Box#1 act as the sources a well as Reservoirs. Box#0 is the Sink. In figure 5(a) all boxes, except Box#0 have equal amount of material to begin with. Figure 5(a) shows the probability distribution $p(\zeta)$ of all the boxes and 5(b) has the $p(\zeta)\ln(1/p(\zeta))$ versus $\zeta$. The emptying profile of the boxes is such that for $\zeta \geq 5$,

Box#0 has higher probability than any other box and by $\zeta \sim 10$, it is about 0.8 while all others are <0.1. The uncertainity $\ln(1/p(\zeta))$ is once again amplified at low $p(\zeta)$. This was highlighted in figure 1(b), (c), (e) and (f) for the case of 2-Boxes. It can be seen in figure 5(b) where self-information or $p(\zeta)\ln(1/p(\zeta))$, of Boxe#5 to #1 contibute significant information even for $\zeta > 10$ where the material transfer probability is very low. The ratio of information generated by the system of material sharing boxes is $I_0 : \sum_1^5 I_b$=1:4.02. It is the entropic cost of 6-Box model that is higher than the corresponding ratio for the 3-Box case. Higher number of boxes imply larger amount of information generation and consequently the higher the entropic cost.

In the 6-Box configuration, the quantity of material contained in each box is varied from equal (Figure 5(a)) to unequal (Figure 5(b)) in the ratio 1:0.8:0.6:0.4:0.2 and simulated with the probabilities $p(\zeta)$ and entropies $p(\zeta)\ln(1/p(\zeta))$, as a function of $\zeta$ in figure 5(c) and (d). Figure 5(c) has the emptying and filing probability $p(\zeta)$ profiles of the six boxes. Unlike figure 5(a), the probability distribution functions have different initial values at $\zeta = 0$ due to unequal starting levels. The net effect of material accumulation, as a function of $\zeta$, can be seen in 5(c) in the build-up of the probability distribution function $p_1(\zeta)$ of Box#1. It serves as the Reservoir for Box#0. The profiles of the two probability distributions $p_1(\zeta)$ and $p_0(\zeta)$ is such that $p_0(\zeta) = 0$ at $\zeta = 0$ as the starting point of the distribution when the Box#0 acting as the Sink starts to fill up. Figure 5(d) shows the entropic profile $p(\zeta)\ln(1/p(\zeta))$ versus $\zeta$ for all six boxes. The entropic cost of the whole process of the transfer of material from five boxes to the Sink (Box#0) is the ratio $I_0 : \sum_1^5 I_b$=1:4.37, slightly higher than the case of equal amount of material in figure 5.

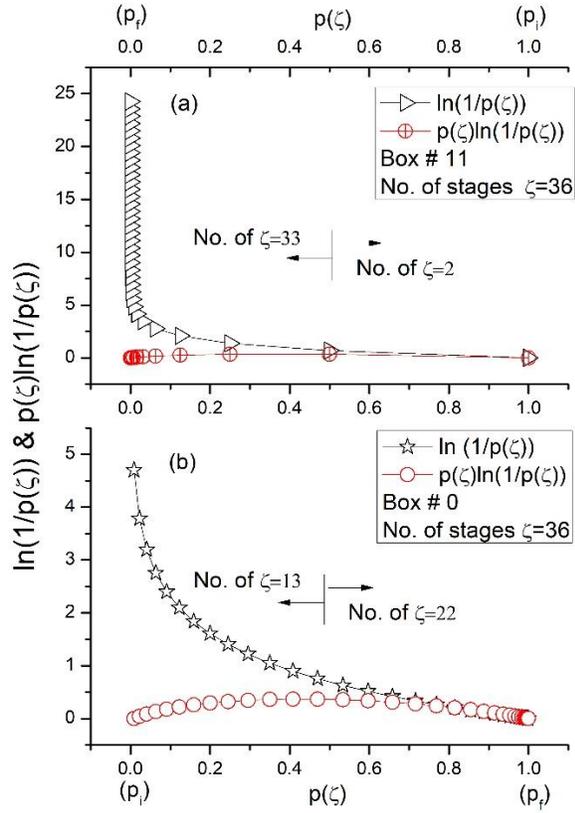

FIG. 6. $\ln(1/p(\zeta))$ and $p(\zeta)\ln(1/p(\zeta))$ are plotted against the probability $p(\zeta)$ for the first Box#5 in 6(a) and Box#0 in 6(b).

As mentioned earlier, increasing the number of boxes, or the stages of transfer of material, increases the entropic cost of information transfer. In physical terms, higher entropic cost implies increased temporal and spatial interactions amongst the constituents that participate in the transfer. The three ingredients of the information, the probability distribution function $p(\zeta)$ the uncertainity $\ln(1/p(\zeta))$, and self-information $p(\zeta)\ln(1/p(\zeta))$, can be best demonstrated when plotted in a single format. Figure 6 has the functions $\ln(1/p(\zeta))$ and $p(\zeta)\ln(1/p(\zeta))$ plotted against the probability $p(\zeta)$. These are done for Box#5 in 6(a) and Box#0 in 6(b). The uncertainty or the 'surprise' around $\zeta \to 0$ is much higher for the emptying box in 6(a) as compared with that of the Sink (Box#0) in 6(b). It is due to the nature of the information transfer. Box#5 in figure 6(a) loses half of its material in the first step. The Box#0 gains the material and the associated information gradually and in multiple steps. The two graphs in 6(a) and 6(b) contain widely different information profiles; one empties $p_5(\zeta) = 1 \rightarrow 0$ and the other fills up $p_0(\zeta) = 0 \rightarrow 1$.

Figure 7 displays fractal dimensional analysis based on information-theoretic entropy $I_x$ denoted in equation (5) as $d_I^x$. It acts as a parameter that identifies the spatial and temporal characteristics of the dissipative structures. In the case of the Boxes, $d_I^x \equiv d_I^{Box\#}$ is plotted for each Box of the configurations discussed earlier in the preceding sections in Figure 7. The figure includes the four sets of boxes that include the largest number of boxes considered in this presentation i.e., 11-Box configuration. As mentioned earlier, the number of boxes i.e. six and eleven have been chosen so that comparison with the physical situation of self-organization among ensembles of carbon cages can be made with the idealized Box-based SRS model. The fractal dimension of the starting Box or the Source is generally lower than those of the succeeding stages except the 2-Box setup. The 2-Box is a special case where there is no reservoir between the Source and the Sink. In such a situation, the emptying box will always generate higher entropy due to the nature of emptying displayed by the lower probabilities as demonstrated in figure 1(b), 1(e) and 6(a). Interestingly, the Sink which receives all the material the whole system generates, through the Reservoir, has lower dimension for 2-, 3- and 5-Box configurations. Only for the 11-Boxes model, $d_I^{Box\#0} \gtrsim d_I^{Box\#1}$. Therefore, the fractal dimension is not a measure of the material content of Sink. Through $d_I^{Box\#}$ all constituents of the SRS model display the interconnected information about the dynamical state of the emerging structures.

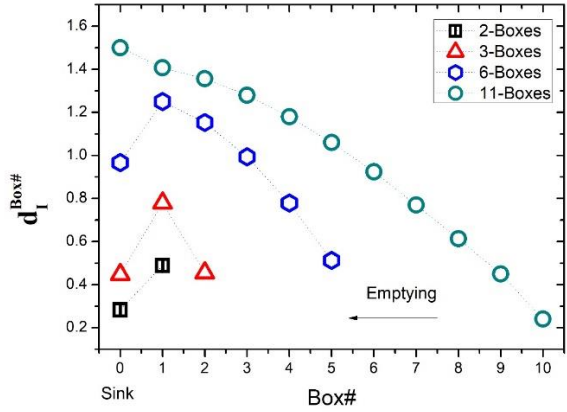

FIG. 7. The fractal dimension $d_I^{Box\#}$ of each box is shown for the 2-, 3-, 6- and 11-Box models. Box#0 is the Sink of all Box configurations. All boxes empty their content or share the information with the next-lower box; the emptying process is indicated by the arrow.

# VI. SUMMARY OF THE SRS MODEL OF DISSIPATIVE STRUCTURES

The basic features of the Source-Reservoir-Sink model of dissipative structures are derived from the results of the preceding discussion. These are

## A. The probability paradigm

All processes and activities relating to the sharing and transfer of material or information can be adequately describaed by the probability distribution functions at every step and stage of the dissipative process. The probability distribution function $p_x(\zeta)$ should describe the properties of the variable *x* at every stage of evolution or the transition. The measure of transition stages $\zeta$ may be externally controlled, as was done for the controllable valves of the Box-system, or these could be defined by the physical nature of the transition, as will be shown in the next section where fragmentation process of fullerenes determines the fragmentation steps or stages.

## B. Information sharing between the Source, Reservoir and the Sink

The sharing and transfer of information among the system constituents-the Source, Reservoir and the Sink, need to be described as a whole by providing the details of the role of each sub-component. The model is based on the rates of the transfer of material and the associated information in an inter-Box information transfer. Except the Sink, all others eventually empty out. By controlling the flow rates one can establish the conditions that may simulate a dissipative system that is driven between different states. The SRS-Box model, for the inanimate material transfer can generate information based on the appropriately derived probability distribution functions, Kolmogrov uncertainity and Shannon entropy. The measures of identifying the end-directed evolution of a dissipative structures are directly linked with the evaluation of the information-theoretic entropy.

## C. The Reservoir of SRS model and the Channel of information theory

The Reservoir of SRS model is analogous to the Channel of Shannon's information transfer network, but with significant differences. (i) The Reservoir may have memory. It can store information to deliver later at variable rates to the Sink or the receiver. The memory could naturally emerge from the different rates of arrival and transmission/delivery of information. (ii) In certain cases, the Reservoir may interact with its own compartments. This may happen in a multiple-source model where more than one sources are interchanging information at various rates. Such conditions have been discussed in the case of 6- and 11-Box scenarios. (iii) It can modify, manipulate or completetly transform the information provided by the Source (or Sources).

It is this capability of the Reservoir that transforms an SRS model into a dissipative dynamical system.

## D. The entropic cost

The net outcome of the SRS model is the information obtained by the Sink as $I_{Sink}$. The ratio of $I_{Sink}$ and the total information generated by the system $\sum_b I_b = I_{Source} + I_{Reservoir} + I_{Sink}$ is defined as the entropic cost $i_b^{\Sigma} = \sum_b I_b / I_{Sink}$. It has been demonstrated earlier and shown in Table 2 that this cost is higher for more effective, more manipulative and end-directed dissipative dynamical systems. For real physical systems, the entropic cost obtained by the technique described in the previous sections, is the minimum for the respective systems. Actual entropic cost is likely to be much higher due to multiple information exchanges between various constituents. In situation where more than one outcomes are possible, then the cost may vary from system to sytem and the ambient conditions of multiple trasitions.

## E. Fractal or the Information dimension

Fractal dimension $d_I^{Box\#}$ for the emptying, storing or filling Boxes is calculated according to equation (4) using the accumulated information for each box. The information $I_x$ for each box increases with the size of the ensemble of boxes. This trend is a natural consequence of the availability of the increasing numbers of the constituents. The ensemble with successively increasing number of boxes generates larger, more varied information content.

**Table II.** Tabulation of entropic cost $i_b^\Sigma = \sum_b I_b / I_{Sink}$ as an impotant information-theoretic parameter of the end-directed evolution of the dissipative multi-Box model is shown for the four box configurations.

|  | 2-Boxes | 3-Boxes | 6-Boxes | 11-Boxes |
|---|---|---|---|---|
| $\sum_b I_b$ | 2.187 | 5.12 | 16.50 | 38.61 |
| $I_{Sink}$ | 0.802 | 1.36 | 3.074 | 5.38 |
| $i_b^\Sigma$ | 2.73 | 3.76 | 5.37 | 7.17 |

## VII. THE CASE OF THE EMERGENCE OF THE BUCKYBALL-$C_{60}$

We apply the SRS model to the case of the emergence of Buckyball with icosahedral symmetry out of the fragmenting, reforming ensembles of fullerene cages with larger number densities but lower symmetries. All of the carbon's closed cages or fullerenes have the unique distinction of cage closure, exactly twelve pentagons among variable number of hexagons and 2-D curvature-related steric strain. The larger cages have higher number of isomers. Fullerenes have the ability for mutual transformation. Fullerenes cycle through cage-to-cage transformations during their formative and fragmentation stages. For optimized cage-transformation conditions, icosahedral $C_{60}$ is the sole survivor. Fullerenes' self-organization in hot carbon vapor is treated here as a dissipative dynamical system whose configuration changes with time. We evaluate the nonlinear interactions among the dynamical system's constituents and their mutual interdependence. Fragmenting, re-forming fullerenes and an evolving gas of $C_2$ are the constituents. Shannon entropies of the constituents are calculated for the iterations of the mapping cage→cage+$C_2$. From the entropic profiles of the transformation of entire fullerene ensembles into $C_{60}$ and $C_2$, the fractal dimensions of the fragmenting and evolving fullerenes are evaluated. Our model describes the conditions for the emergence of $C_{60}$ as the end-directed evolution of self-organizing dissipative dynamical systems of the transforming cages.

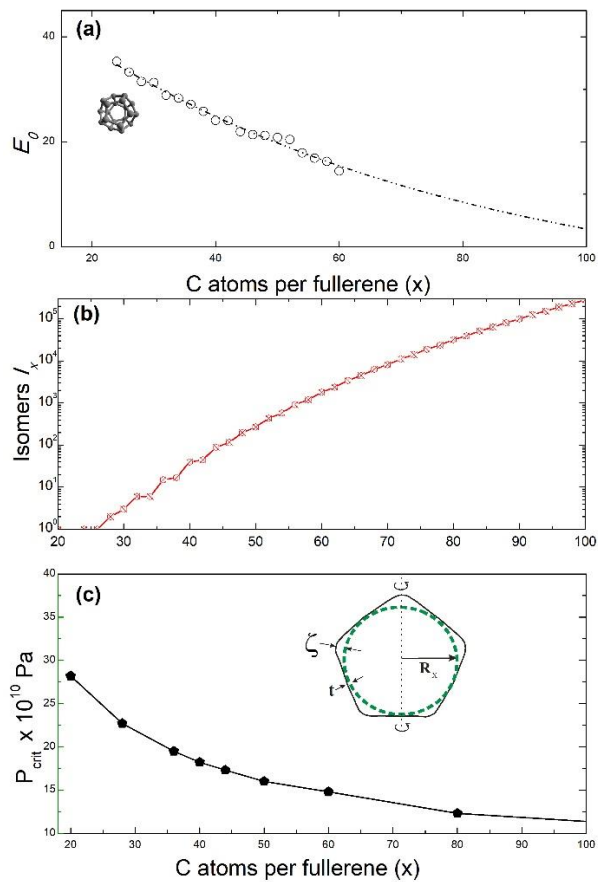

FIG. 8. (a) The carbon cage closure is favoured with decreasing energy of formation for larger cages, (b) fullerenes' isomer density increases for larger cages. (c) The fragmentation forces related with cage imperfections are integral component of each fullerene with pentagon-related imperfections.

The driving forces of fragmentation are related to the inherent, structural instability due to the pentagon-related internal stresses and the degeneracy pressures of the Fermi gas of pi electrons [43]. Hot carbon vapor in an appropriately confined environment can generate fullerenes with relative densities that are determined by their respective heats of formation and isomeric abundance. Various mechanisms for the emergence of $C_{60}$ out of the disorder of the hot C-vapor have been proposed ever since the discovery of Buckyball. Here we treat the grand canonical ensemble of the forming and fragmenting fullerenes as a dissipative dynamical system. By treating fullerenes as 3D rotors, a thermodynamic model was developed earlier with evaluation of thermal entropy of cage-to-cage transformations [37]. The thermodynamic entropic arguments have recently been re-evaluated by the Shannon entropic considerations. Probability is calculated for all steps of the sequences of dynamic transformation of fullerenes. The information or Shannon entropy is obtained by the sum of $p_x(\zeta)\ln(1/p_x(\zeta))$ over all fragmenting stages, describes the dissipative structure of all constituents of the dynamical systems.

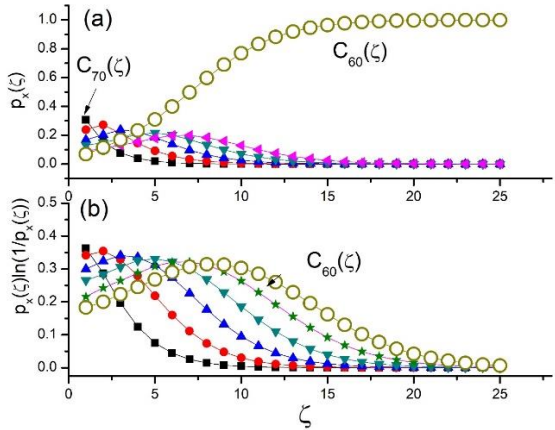

FIG. 9. The fragmentation profile of an ensemble of fullerenes $\sum_{60}^{70} C_x$. Fig. 1(a) and (b) show the profile of the cage transformations. For each cage with *x*-carbon atoms, $p_x(\zeta)\ln(1/p_x(\zeta))$ is computed and plotted at every fragmentation step $\zeta$. At fragmentation step $\zeta=0$, all cages start with their respective isomeric densities. Gradual build-up of $C_{60}$ is visible from 1(a) and 1(b).

Each fullerene $C_x$ is mapped onto its fragmented cage $C_{x-2}$ and a $C_2$ molecule as $f: C_x \to C_{x-2} + C_2$. The iterations of cage→cage transformations are carried out for four such self-organizing, fragmenting dynamical systems designated as $\sum_{60}^{70} C_x$, $\sum_{60}^{80} C_x$, $\sum_{60}^{90} C_x$ and $\sum_{60}^{100} C_x$ of fullerenes have been investigated. The range of fullerenes in each ensemble is from the lower to the higher limit of summation. In Figure 9 the results of iterations for the ensemble of six fullerenes of the ensemble $\sum_{60}^{70} C_x$ are shown for each fragmenting step $\zeta$. Figure 9(a) shows the normalized probability $p_x(\zeta)$ evaluated for each fullerene $C_x$ of the ensembles, at each fragmenting stage $\zeta$. Figure 9(b) has $p_x(\zeta)\ln(1/p_x(\zeta))$ of all the cage-transformations calculated and plotted. It shows the entropic profile of all cage-to-cage transformations. $C_{60}$ can be seen emerging out in the probability as well as the entropic graphs. The emergence of $C_{60}$ is the end-directed result of cumulative fragmentations of all fullerenes into the next smaller one. All fullerene profiles, except that of $C_{60}$, self-organize into the entropic curve of $C_{60}$ as the net outcome of the dissipative, nonlinear dynamical system of ensembles of fragmenting, non-icosahedral fullerenes.

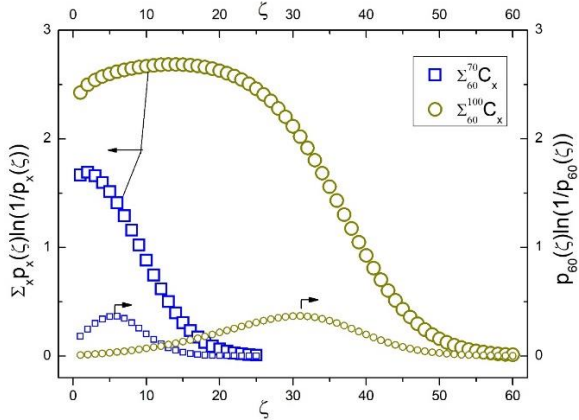

FIG. 10. The phase profile of all fullerenes of the two ensembles $\sum_{60}^{100} C_x$ and $\sum_{60}^{70} C_x$ are plotted as a function of $\zeta$. The corresponding entropic profile of the emerging $C_{60}$ is also shown as the sink or the end-directed product. The left vertical axis shows the sum of entropies of all cages at each $\zeta$ for the whole ensemble and the right vertical axis has the emerging entropic values for $C_{60}$.

The phase profile of the shrinking and disappearing fullerenes as a function of the fragmenting steps are shown for the two ensembles $\sum_{60}^{100} C_x$ and $\sum_{60}^{70} C_x$ in figure 10. The sum of instantaneous entropies of all the cages from $x=60$ to 70 and 100, are shown at each fragmenting step $\zeta$. Each point represents the state of the entire ensemble. The entropic profiles for $C_{60}$ in the two ensembles are also shown. We have included all cage transformations of the type $C_x \to C_{x-2} + C_2 \to C_{x-4} + C_2 \to \cdots C_{60} + C_2$. The sum over all $x$ and $\zeta$ of $p_x(\zeta)\ln(1/p_x(\zeta))$ is the entropic cost of the dynamical transition of the entire ensemble of fullerenes $\sum_x C_x(\zeta)$ into the two gases of $C_{60}$ and $C_2$. The phase trajectories of transformation of the ensembles represent the dynamic profile of the self-organizing fullerenes. Due to the nature of assumptions, the total number of cages remains constant in $cage \to cage$ transformations, only that of $C_2$ increases with each fragmentation..

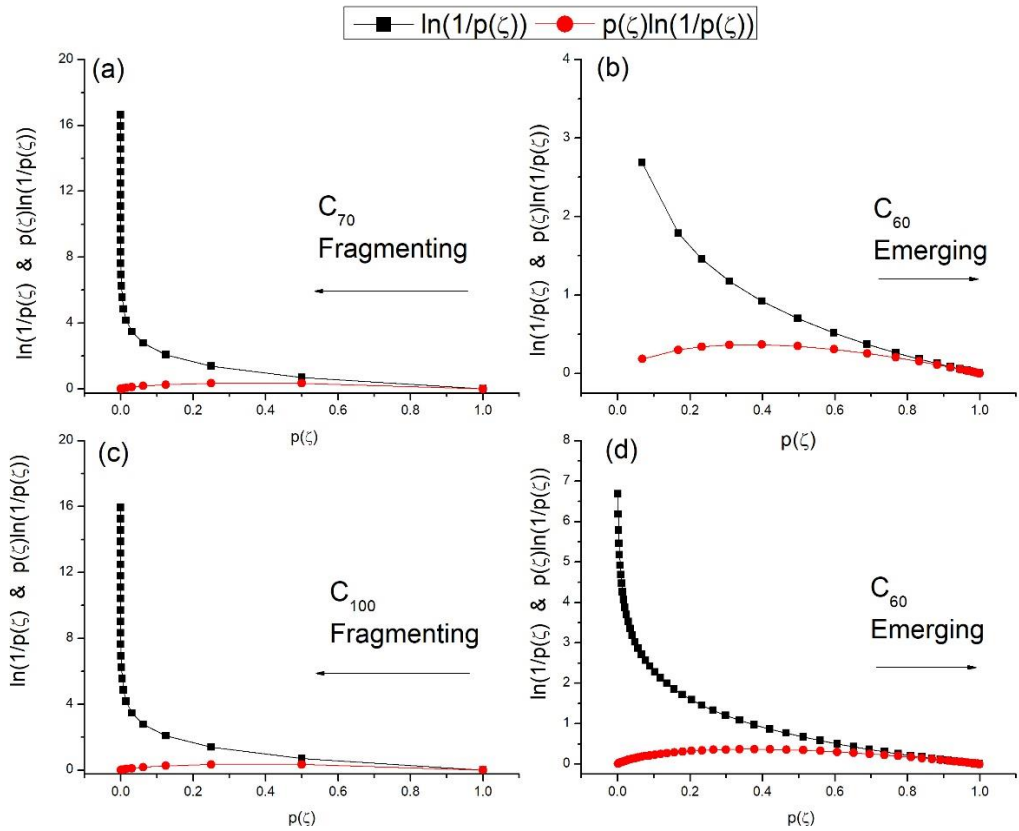

FIG. 11. The information theoretic parameters $ln(1/p_x(\zeta))$ and $p_x(\zeta)\ln(1/p_x(\zeta))$ for the two ensembles are plotted for the two fullerenes; the largest and the emerging one. For $\sum_{60}^{70} C_x$ the fragmenting $C_{70}$ is shown in (a) and for $C_{60}$ in (b); one fragmenting and the other is the emerging fullerene. (c) is the fragmenting profile of $C_{100}$ and (d) for $C_{60}$ for the ensemble $\sum_{60}^{100} C_x$.

Figure 11 is the real physical world application of the SRS model developed for the interconnected, information generating and sharing boxes. The two sets of fragmenting fullerenes belonging to the ensembles $\sum_{60}^{70} C_x$ and $\sum_{60}^{100} C_x$ operate as the information-sharing Boxes. We have used the same rates of fragmentation i.e., the @1/2 which implies half of all the fullerens fragment and transform into the next smaller ones as indicated in the mapping scheme $C_x \rightarrow C_{x-2} + C_2 \rightarrow C_{x-4} + C_2 \rightarrow \cdots C_{60} + C_2$. The cumulative results were shown in figures 9 and 10. Here the functions $\ln(1/p_x(\zeta))$ and $p_x(\zeta)\ln(1/p_x(\zeta))$ for the initial two fullerenes $C_{70}$ and $C_{100}$ are presented in figure 11(a) and (c). These are the profiles of the fragmenting fullerenes. The emerging cage is $C_{60}$ that is shown in the two sequences of transformations in figure 11(b) and (d).

**Table III.** The Shannon entropies or Information for all of the cages are tabulated for four ensembles of fullerenes. These are shown as $\sum I_x \equiv \sum_\zeta \sum_x p_x(\zeta)\ln(1/p_x(\zeta))$, $I_{60} = \sum_\zeta p_{60}(\zeta)\ln(1/p_{60}(\zeta))$, the entropic cost $i_{60}^\Sigma$ and the fractal dimension $d_I^{60}$ of the emergent structure $C_{60}$.

|  | $\sum_{60}^{70} C_x$ | $\sum_{60}^{70} C_x$ | $\sum_{60}^{90} C_x$ | $\sum_{60}^{100} C_x$ |
|---|---|---|---|---|
| $\Sigma I_x$ | 19.94 | 45.06 | 71.88 | 103.97 |
| $I_{60}$ | 3.4 | 5.73 | 7.54 | 8.83 |
| $i_{60}^\Sigma$ | 5.86 | 7.86 | 9.53 | 11.77 |
| $d_I^{60}$ | 1.13 | 1.61 | 1.9 | 2.1 |

Information emerges as the crucial parameter that defines the phase space of a dissipative dynamical system and is employed in this communication to evaluate fractal dimensions of its constituents. The data for the four dynamical systems are collected in Table III. It has the collective Shannon entropies or the Information for all of the ensembles and the evolving gas of $C_{60}$s as $\sum I_x \equiv \sum_\zeta \sum_x p_x(\zeta)\ln(1/p_x(\zeta))$ and $I_{60} = \sum_\zeta p_{60}(\zeta)\ln(1/p_{60}(\zeta))$. The fractal dimension $d_I^{60}$ of the accumulating gases of $C_{60}$ has been calculated for the two sets of fullernes belonging to ensembles $\sum_{60}^{70} C_x$ and $\sum_{60}^{100} C_x$. Figure 12 is derived from the the information data for the fragmenting, constituent fullerenes of the ensembles and the emerging $C_{60}$.

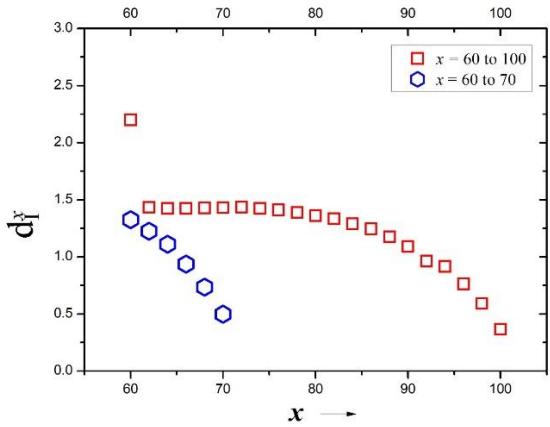

FIG. 12. Fractal dimension $d_I^x$ of all fullerenes that are constituents of the two ensembles $\sum_{60}^{70} C_x$ and $\sum_{60}^{100} C_x$ are plotted against the number of carbon atoms $x$.

In Figure 12 fractal dimensions of fragmenting fullerenes of the two ensmbles $\sum_{60}^{70} C_x$ and $\sum_{60}^{100} C_x$ are shown. These are calculated from the information generated during the complete transition where the entrie ensemble has coalesced into the Buckyball. There is the trend of increasing information generation by the successive smaller cages. $C_{70}$ and $C_{100}$ generated the lowest information and hence the smallest fractal dimension. This trend was demonstrated by the multi-Box examples of the SRS model in Figure 7. Fractal dimension is being presented here as the measure of the emergence in the dynamical system of the cage transformation sequences.

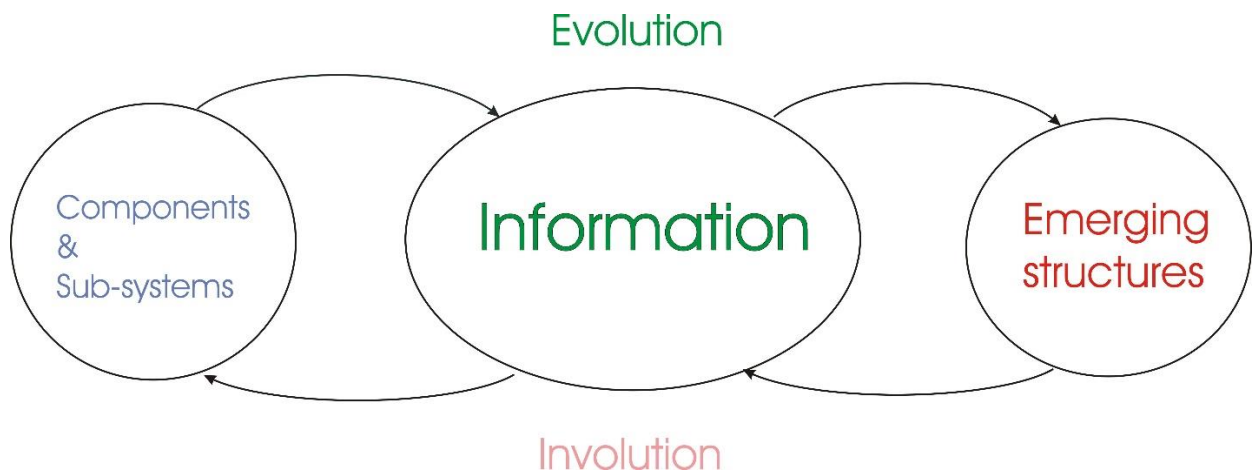

FIG. 13. A generic diagram of the evolution of the emerging structures in dynamical dissipative environment from sub-systems or components. Information is shown to play the crucial role in the emergence of complex structures.

The main objective of Figure 13 is to clarify some of the concepts that were introduced in this communication. It is also intended to suggest two basic points. (i) The first deals with the complex structures that emerge due to the self-organizing sub-structures or the components as the basic ingrdients. Self-organization requires energy input and appropriate environment. As we are dealing with carbon cages, therefore we assume this environment be the carbon vapor generated by laser ablation of graphite in vacuum. The constituents of this vapor will be the components and sub-systems of Figure 13. This stage generates carbon cages of all shapes, sizes, isomeric variations and with pentagonal curvature-induced structural imperfections. We had reported earlier experimental results of such an evolution of carbon cages under high energy, heavy ionic irradiations of amorphous graphite where fullerenes ensembles consisted of the cages containing from ~30 to 90 carbon atoms [11-13]. These experiments were at room temperature where the sputtered atoms and molecules assembled into cages. These were not further transformed due to the lack of appropriate physical conditions and environment. It could be considered the 'stunted' growth stage of the pre-self-organizing ensembles of carbon cages. Another classic example is that of the pre-Buckyball discovery experiments of Rohlfing et al [44]. They had ablated graphite with laser and mass analyzed the emerging carbon 'clusters'. They did not recognize the importance of $C_x > C_{30}$ as the closed cages-the fullerenes. Nor did they provide the environment of self-organization. Kroto and Smalley recognized the importance of cage-transformations as a result of the variations in the design of the extractor cone, pulses of He and other physical parameters [30]. In the context of our previous discussion, Rohlfing experiment falls in the category of the 'stunted growth' stage, like our experiments [11-13]. (ii) The second point deals with the next stage where these initially formed cages are the components and the sub-systems. The self-organization requires that the hot carbon atoms, molecules and cage-containing vapor continue the collision sequences of cage-cage, molecules-cage, molecules-molecules and of course, the collisions of all species with the cooling gas (He or Ar) and with the confining walls of the expansion chamber. It is here that the second stage of cage-transformations, discussed in detail earlier, begins. The cumulative information of the cage transformation processes provides clues to the evolution of complex structures. Self-organization is not a one-way process of evolution; involution also operates on the system such that the reverse sequences may also be operating to optimize the evolving structures like $C_2 + C_x \rightarrow C_{x+2}$. The figure schematically describes these stages of involution and evolution.

## VIII. CONCLUSIONS

An information-theoretic model has been presented that has its general features and as a special case it can describe the dissipative dynamical systems of ensembles of fragmenting, self-organizing fullerene cages. Probabilities of the changing relative densities of the cages for each fragmentation stage are used to evaluate information or Shannon entropies for every cage. Fractal dimension of all the constituents of the dynamical system are calculated from their respective Shannon entropies. $C_{60}$ emerges as the end-directed evolution of the dynamical system of four different ensembles of fullerenes. The detailed description of the emergence of $C_{60}$ is provided by the dynamic variations of the phase space of all the constituents of the ensembles of fragmenting fullerenes and the evolving gases of $C_{60}$ and $C_2$. The dissipative structure of the information-generating, fragmenting cages gives us a perspective to evaluate the self-organizational behavior. Summary of the SRS and the Box model applied to the self-organizing fullerenes is following:

1. A model for the self-organizing fullerenes in hot carbon vapor environment where cage formation and fragmentation takes place, is developed. Self-organizing fullerene ensembles are modeled as dissipative dynamical systems.

2. Starting with densities in the ratio of the isomeric possibilities of each of the fullerene, each cage is subjected to the internal cage-fragmentation forces, the starting probabilities are evaluated.

3. The larger cages transform into the successive smaller ones. From the density variation of cages, the probabilities of the entire cage-to-cage transformations are calculated.

4. The Information or Shannon entropy summed over the entropic profiles of all fragmenting stages is first tabulated and then it is used to calculate fractal dimension of all constituents of the dynamical systems.

5. We have shown that the dissipative dynamical system of fragmenting and transforming cages has an end-firected evolution towards $C_{60}$. It acts as the sink whose basin is the entire ensemble.

6. Information is the crucial link between the phase space of a dissipative dynamical system and fractal dimensions of its constituents. It helps us to identify the processes of self-organization.

Figure 14 is the graphical summary of the SRS model applied to the self-organizing fullerenes. An ensemble $\sum_{60}^{80} C_x$ consisting of 11 fullerenes $C_{60}$, $C_{62}$, $C_{64}$, $C_{66}$, $C_{68}$, $C_{70}$, $C_{72}$, $C_{74}$, $C_{76}$, $C_{78}$ and $C_{80}$ transform into $C_{60}$ via $\sum_{60}^{80} C_x \rightarrow \sum C_{60} + \sum C_2$. Feedback mechanisms are highlighted. Involution and feedback of material and information are the essential elements of self-organization.

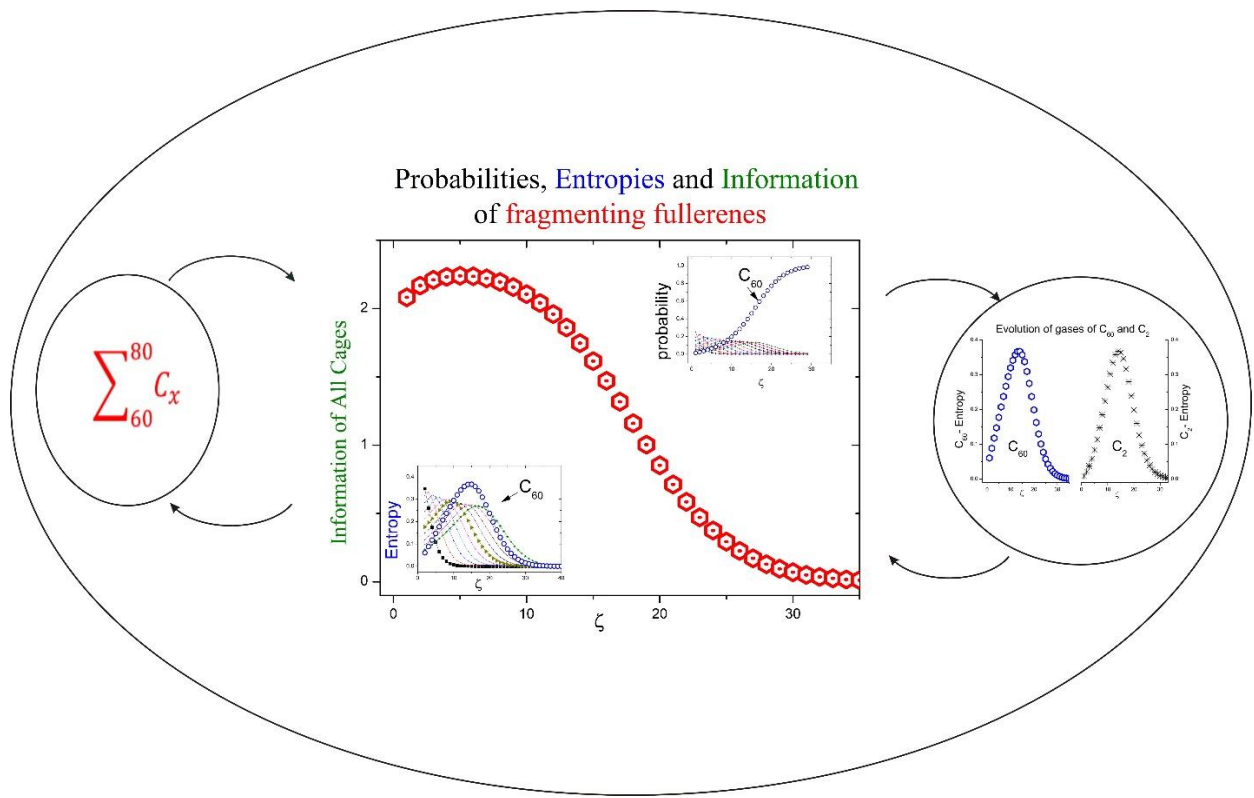

FIG. 14. The ellipse of information: The emergence of $C_{60}$ out of the ensemble of fragmenting and reforming fullerenes is shown as the end-directed evolution of the dynamical system $\sum_{60}^{80} C_x \rightarrow \sum C_{60}$. It describes graphically the generation of information, instantaneous and total by the ensemble of fragmenting fullerenes. The information is in the form of probabilities and entropies. A typical ensemble of 11 fullerenes from $C_{60}$ to $C_{80}$ represented by $\sum_{60}^{80} C_x$ and shown in the extreme left ellipse, is chosen as the example. The number densities of ensemble's fullerenes are proportional to their isomeric densities. Each, non-icosahedral fullerene, is subjected to the internal cage-fragmentation forces described in Figure 8. The ensemble is subjected to the hot sooty environment of the condensing carbon vapor that results in the aftermath of laser ablative pulses. The model assumes fragmentation of half of all fullerenes at each fragmentation stage $\zeta$. The larger fragmenting cages populate the smaller cage densities as described in section V and displayed graphically in Figure 7.

## Acknowledgements

The author is grateful for discussions on various aspects of the model with colleagues S. Javeed, S, Zeeshan, M. Yousuf, M. A. Abbas, A. Afshan, K. Yaqub. Gratitude is expressed to PIEAS for giving the course Fractals and the Irradiated Solids in the spring of 2018 that provided opportunites for extended discussions and debate with young scholars and colleagues.